\documentclass[12pt]{article}
\usepackage{graphicx}
\usepackage{epsfig}

\usepackage{amssymb}

\begin{document}

\title{Impact of scaling range on the effectiveness of detrending methods}
\author{Dariusz Grech$^{(a)}$\footnote{dgrech@ift.uni.wroc.pl} and Zygmunt Mazur$^{(b)}$\footnote{zmazur@ifd.uni.wroc.pl}}
\date{}

\maketitle
\begin{center}
\begin{flushleft}
(a) Institute of Theoretical Physics, University of Wroc{\l}aw, Pl. M. Borna 9,\\ PL-50-204 Wroc{\l}aw, Poland

(b) Institute of Experimental Physics, University of Wroc{\l}aw, Pl. M. Borna 9, \\PL-50-204 Wroc{\l}aw, Poland
\end{flushleft}
\end{center}

\hspace{1.5 cm}

\begin{abstract}
We make the comparative study of scaling range properties  for detrended fluctuation analysis (DFA), detrended moving average analysis (DMA) and recently proposed new technique called modified detrended moving average analysis (MDMA). Basic properties of scaling ranges for these techniques are reviewed. The efficiency and exactness of all three methods towards proper determination of scaling Hurst exponent $H$ is discussed, particularly for short series of uncorrelated and persistent data.
\end{abstract}
$$
$$

\textbf{PACS:} 05.45.Tp, 89.75.Da, 02.60.-x, 89.20.-a
$$
$$

 In last years much effort was put into precise analysis of scaling range for power laws used to classify long-term memory properties in complex systems and in time series [1--7].
This effect in stationary time series $x_t$, $(t=1,...,L)$, is usually attacked  with two point autocorrelation function $C_s=\langle\Delta x_t \Delta x_{t+s}\rangle$, where $\langle\, \rangle$ is the average taken from all data in the signal with increments $\Delta x_t = x_{t+1}-x_t$. The scaling law is proven \cite{gamma_1,gamma_2}
\begin{equation}
C_s\simeq (2-\gamma)(1-\gamma)s^{-\gamma},
\end{equation}
where $s$ is the time lag and the autocorrelation scaling exponent $\gamma$ $(0\leq \gamma \leq 1)$ describes the level of long-term memory in a signal.
 The two edge values $\gamma=0$ and $\gamma=1$ are related respectively to fully correlated or uncorrelated (integer Brownian motion) data, leaving space for other fractional persistent Brownian motion ($0< \gamma < 1$) in between.

 However, a direct calculation of correlation functions and $\gamma$ exponent may suffer from problems connected with noise present in time series or possible non-stationarities in data (and thus locally changing the quality of power law in Eq.(1)). Therefore, it is recommended to
reduce these effects not calculating $\gamma$ directly, but
 studying instead the "integrated profile" of the data, i.e., the random walk $x_t$ instead of $\Delta x_t$ behavior. In the latter case
 the scaling Hurst exponent $H$ of the series \cite{Hurst1,Hurst2} is measured. Traditionally, $H$ is defined as the exponent of the power law relation
\begin{equation}
var(x_t) = \langle x_t^2\rangle-\langle x_t\rangle^2\simeq t^{2H}
\end{equation}
where $var(·)$ is the variance calculated in a time window of length $t$.
 All methods calculating $H$ are useful in analysis of long-term memory properties in data since there exists a simple formula linking $\gamma$ and $H$ exponent for large $t$ \cite{rel_gamma_H}
 \begin{equation}
H=1-\frac{\gamma}{2}
\end{equation}
 Furthermore, to avoid an artificial bias in $x_t$ data caused by the presence of trend influencing the final outcome for $H$, the so called detrending procedure is recommended. Two efficient major techniques were proposed in literature to do so: detrended fluctuation analysis (DFA) \cite{DFA_1,DFA_2} and detrended moving average analysis (DMA) \cite{DMA_1,DMA_2,DMA_3,DMA_4} with variety of their 'clones' applicable also for multifractals \cite{multi}, where two point autocorrelation functions are not sufficient to describe the variety of autocorrelation properties in data. Recently DMA was generalized to its modified version called MDMA \cite{MDMA} where the statistics of data points used to calculate the trend in signal is more balanced than in case of DMA.

 The scaling law from Eq.(1) has been built into DFA in a form of power law
\begin{equation}
F^2(\tau)\simeq \tau^{2H}
\end{equation}
where $F(\tau)$ is the averaged fluctuation of the signal around its local trend in  time windows of fixed length $\tau$.  To be more precise:
\begin{equation}
F^2(\tau)=\frac{1}{2N}\sum^{2N}_{k=1} \hat{F}^2(\tau,k)
\end{equation}
where
\begin{equation}
\hat{F}^2(\tau,k)=\frac{1}{\tau}\sum^{\tau}_{j=1}\left\{x_{(k-1)\tau+j}-P_k(t)\right\}^2
\end{equation}
Here $N=[L/{\tau}]$ stands for the number of non-overlapping boxes obtained after cutting the whole walk $x_t$ ($t=1,...L$) into separate pieces where detrending is performed with a polynomial trend $P_k$ fitted to data in $k$-th window box.

The DMA method serves the similar power law

\begin{equation}
F_{DMA}^2(\lambda)\sim \lambda^{2H}
\end{equation}
but here the fluctuation function (variance) is defined according to

\begin{equation}
F_{DMA}^2(\lambda) =\frac{1}{L-\lambda +1}{\sum_{t=\lambda}^L (x_t - \langle x_t\rangle_{\lambda})^2}
\end{equation}
where $\langle x_t\rangle_{\lambda}$ is the moving average of length $\lambda$ calculated as
\begin{equation}
\langle x_t\rangle_{\lambda}=\frac{1}{\lambda}\sum_{k=i-\lambda +1}^{t} x_k
\end{equation}
and plays the role of a trend.

 The latter method suffers however from diversified statistics of data points contributing to fluctuation function $F_{DMA}^2(\lambda)$, since  only data points $x_t$ with $t\geq\lambda$ can be taken into account for determination of the variance $F_{DMA}^2(\lambda)$. Thus the statistics depends strongly on the chosen length $\lambda$ of the moving average and for particular choice of $\lambda$ and $L$ only $L-\lambda+1$ detrended values contributes to the $F_{DMA}^2(\lambda)$  in the power law of Eq.(7).

This difficulty can be omitted in the modified DMA technique (MDMA) \cite{MDMA}. Its modification is based on the assumption that usually more than $L$ data points actually exist in a real time series one investigates and some amount of data stored before the basic series of length $L$, although not explored, is actually available for study.
The available amount of data can therefore be written as the sequence $\{x_{-\lambda_{max}},...,x_{-2},x_{-1},x_1,x_2,...,x_L\}$, where $\lambda_{max}$ is the maximal scaling range used in particular calculation (determination of $H$). Thus one is able to calculate trends (moving averages) for those data points where DMA procedure with particular choice of $\lambda$ simply fails. To be precise, Eq.(8) is now replaced by
\begin{equation}
F_{MDMA}^2(\lambda) =\frac{1}{L}\sum_{t=1}^L (x_t - \widetilde{\langle x_t\rangle}_{\lambda})^2
\end{equation}
with the moving average of length $\lambda$ calculated for $t\geq\lambda$ according to Eq.(9), while modified for $0<t<\lambda$ to
\begin{equation}
\widetilde{\langle x_t\rangle}_{\lambda}=\frac{1}{\lambda}(\sum_{k=1}^{t} x_k + \sum_{k=t-\lambda}^{-1} x_k)
\end{equation}
where $k<0$ means that summation takes into account additional data points preceding the basic series.

The power law similar to Eq.(7) is still expected where $F_{DMA}^2(\lambda)$ is replaced by $F_{MDMA}^2(\lambda)$, i.e.
\begin{equation}
F_{MDMA}^2(\lambda)\sim \lambda^{2H}
\end{equation}

The precise value of the scaling range for power laws given by Eqs.(4)(7)(12) will be a function of available length of a signal $L$, i.e., $\lambda = \lambda(L)$ and $\tau=\tau(L)$ but obviously they will depend also on the accuracy of fit $R^2$ of the scaling law -- usually represented in log-log scale as a linear regression fit. One expects also that dependence on the persistency level in data may occur. Hence, the general analysis of this problem may be quite complex and should be made step by step with convincing statistics of synthetically generated time series with a priori known autocorrelation properties. The persistent time series can be generated using the Fourier filtering (FFM) algorithm \cite{FFM}. The scaling behavior of two point autocovariance function $C_s=\langle\Delta x_t \Delta x_{t+s}\rangle$ can then be checked qualitatively and quantitatively as in Ref.\cite{PhysAmulti2013}.

The detailed study of dependencies shown in log-log scale in Figs.\,1--2 and parameters extracted from those fits convinces that the final relationship between scaling range $\lambda$, signal length $L$ and the accuracy of fit $R^2$ takes both for DMA and MDMA the power law form

\begin{equation}
\lambda(L, R^2)) = D L^\eta  (1-R^2)^\xi
\end{equation}
where $D$, $\eta$ and $\xi$ depend only on the method (DFA, MDMA) and on the persistency level in data (see Ref. \cite{MDMA} for details). Moreover, the fitted values of these parameters are linear in $\gamma$ exponent in the first approximation.

A similar consideration for DFA leads to plots like in Figs.\,3--4 but, on the contrary,  prepared in linear scale this time. It supports the relationship
\begin{equation}
\lambda(L, R^2) = (AR^2+ A_0)L + B
\end{equation}
where parameters $A$, $A_0$ and $B$ also depend only on data persistency and  are linear with $\gamma$ for a wide range of $0<\gamma<1$  \cite{former_paper}.

However, the knowledge of $\lambda(L,R^2)$ dependence is still not sufficient for practical use, since we do not know whether the scaling exponent $H$ is properly reproduced, even if the scaling law of Eqs.\,(4)(7)(12) is firmly confirmed for given scaling range $\lambda$. Therefore, it is worth discussing the efficiency  of all three methods in precise determination of Hurst exponents when a precisely determined scaling range is taken for calculations.

Many particular approaches can be proposed for a such project. Here we provide some preliminary outcomes obtained for the following questions. How will the outcome (measured value of $H$ exponent in DFA, DMA or MDMA) depend on the chosen scaling range $\lambda$ for synthetic series of data with precisely given autocovariance exponent $\gamma$ as the input? What is the scaling range of all three methods, most effectively reproducing this input value $H_{in}$ at the assumed confidence level (we assumed in this approach $|H-H_{in}|/H_{in}\leq 1\%$)?

The answers can be deduced from plots like in Figs.\,5--9. Not all plots for variety of possible parameters are presented here because they look qualitatively similar. Fig.\,5 shows the reproduced $H$ value as a function of chosen scaling range for three discussed detrending methods and for two distinct lengths of random walk data. Fig.\,6 indicates the same dependence for persistent signal. An answer to second problem stated above is given in Figs.\,7-9.

 One notices from Figs.\,5,\,6 that DFA reproduces $H$ in the most stabile way but simultaneously underestimates its real value for persistent series. We have checked that for $H_{in}\geq 0.7$ the outcome value of Hurst exponent determined within DFA will always lie below $H_{in}$ and this discrepancy grows with scaling range. The MDMA method reproduces input $H$ value less stabile than DFA but offers better performance in retrieving $H$ exponent than DMA. The MDMA  reproduces $H$ exponent value better than DMA and worse than DFA for longer scaling ranges -- independently on persistence level in data. For very short ranges ($\lambda \leq 10^{-1}L$) DMA and MDMA are not distinguishable, but both methods slightly overestimate $H$ value for such scaling range. They underestimate $H$ for $\lambda > 10^{-1}L$ but this underestimation is more gentle in case of MDMA, particularly for uncorrelated or weakly autocorrelated time series.

Most important results can be read from Figs.\,7,\,8. These plots show the actual scaling range $\lambda^*$ for which the real value of $H$ (or $\gamma$) exponent is strictly reproduced. It turns out that for all detrending methods (with an exception for DFA applied to persistent signal) the power law relation seems to be valid

\begin{equation}
\lambda^* = AL^{m}
\end{equation}
where the parameters $A$ and $m$ are found from the linear fit in double log scale and are collected in Table 1 for four  Hurst exponent values $H=0.5$, $H=0.6$, $H=0.7$, $H=0.8$. Table 1 presents also uncertainties of the best  fit estimation  of $A$ and $m$ from linear plots of Fig.\,7 at $1\sigma$ level. It turns out after that such uncertainty leads to reproduction of $H$ at a very demanding level ($\delta H/H \lesssim 1\%$).

These results can be translated into magnitude of percentage \textit{relative} scaling range $\lambda^*/L$ shown in Figs.\,8--9, which retrieves  exactly the input value of Hurst exponent. We see that for short time series ($L<2000$) one needs to take longer scaling range ($10\% - 25\% L$) in case of DMA and MDMA to do so. The DFA is much less demanding here. If $L>2000$ it is sufficient to take even less than $10\% L$ to calculate $H$ exactly.

Concluding, we may say that MDMA is overall somewhere between--it wins over DMA but loses with DFA. Only in the exceptional case of persistent data DFA is a loser, since MDMA and DMA are capable to reproduce correctly the input  $H$ value for short scaling ranges ($\lambda \leq 10^{-1}L$), while DFA method fails to do so.

\begin{table}
\centering
\begin{tabular}{||c||c|c|c||}
\hline
method & DFA & DMA & MDMA\\
\hline
parameter & log A & log A & log A\\
\hline
H=0.5 & $0.240\pm 0.082$ & $0.190\pm 0.051$ & $0.343\pm 0.061$\\
\hline
H=0.6 & $0.982\pm 0.040$ & $0.180\pm 0.055$ & $0.325\pm 0.057$\\
\hline
H=0.7 & X & $0.207\pm 0.046$ & $0.343\pm 0.030$\\
\hline
H=0.8 & X & $0.381\pm 0.042$ & $0.504\pm 0.043$\\
\hline
\hline
parameter & m & m & m\\
\hline
H=0.5 & $0.623\pm 0.023$ & $0.694\pm 0.015$ & $0.661\pm 0.018$\\
\hline
H=0.6 & $0.234\pm 0.012$ & $0.675\pm 0.016$ & $0.642\pm 0.016$\\
\hline
H=0.7 & X & $0.656\pm 0.013$ & $0.625\pm 0.009$\\
\hline
H=0.8 & X & $0.548\pm 0.012$  & $0.518\pm 0.012$\\
\hline
\end{tabular}
\caption{Fit for parameters in Eq.(15) describing scaling range at which the actual input value of scaling exponent $H$ is well reproduced. The cross mark indicate that no solution is available for particular method. The uncertainties shown here come from the best  fit estimation  of $A$ and $m$ in Fig.\,7 at $1\sigma$ level.}
\label{tab1}
\end{table}

\begin{figure}
\begin{center}
{\psfig{file=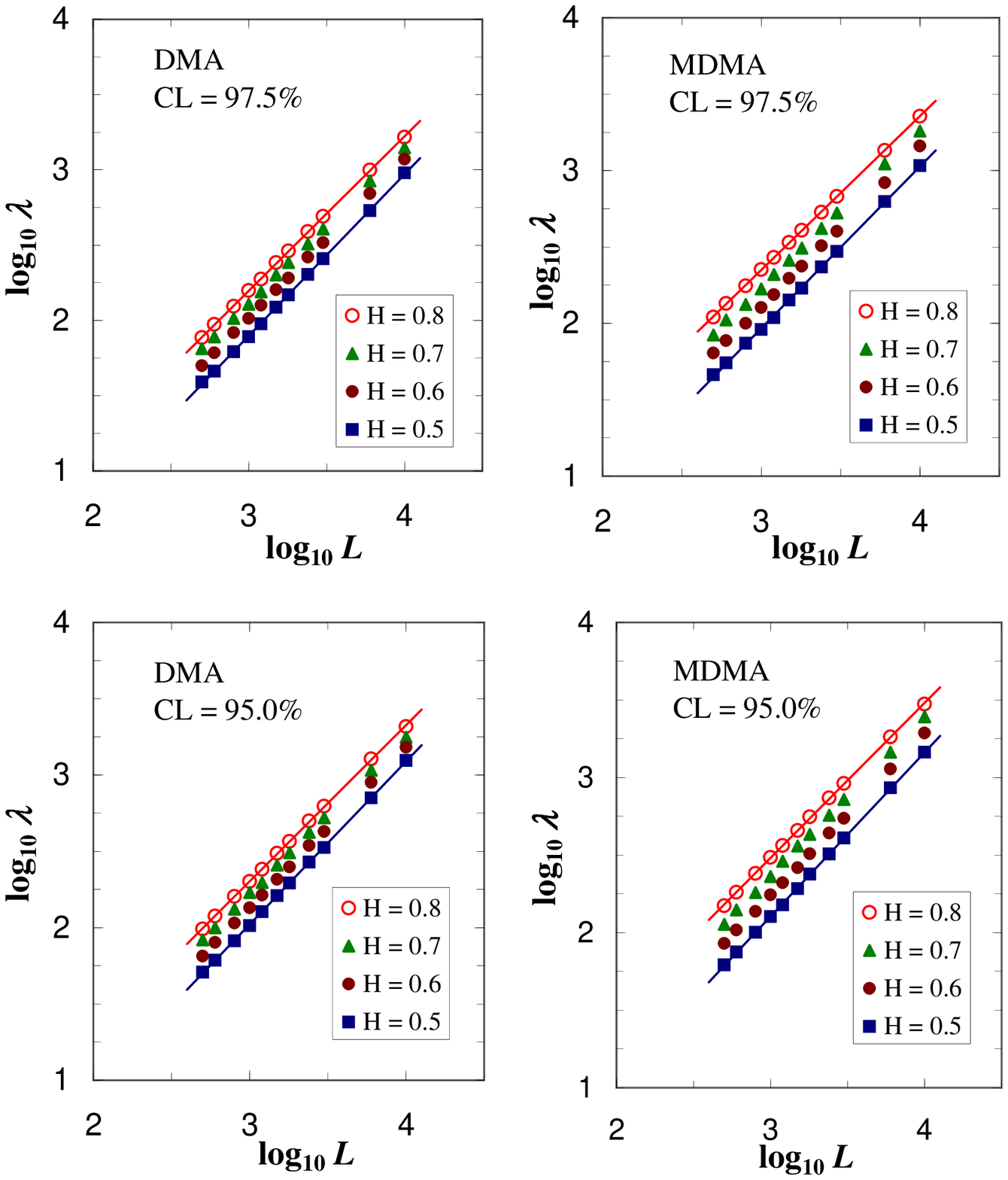,width=15cm}}
\end{center}
\caption{Dependence between scaling range $\lambda$ and length of time series $L$ for various levels of autocorrelation in data (measured by  $H$ exponent). The plots are drawn for particular choice $R^2= 0.98$ for two confidence levels $CL=97.5\%$ and $CL=95\%$. For other $R^2$ values (not shown) they look qualitatively similar. The fitting lines are drawn only for edge values $H=0.5$ and $H=0.8$ to make all remaining dependencies more readable.}
\end{figure}

\begin{figure}
\begin{center}
{\psfig{file=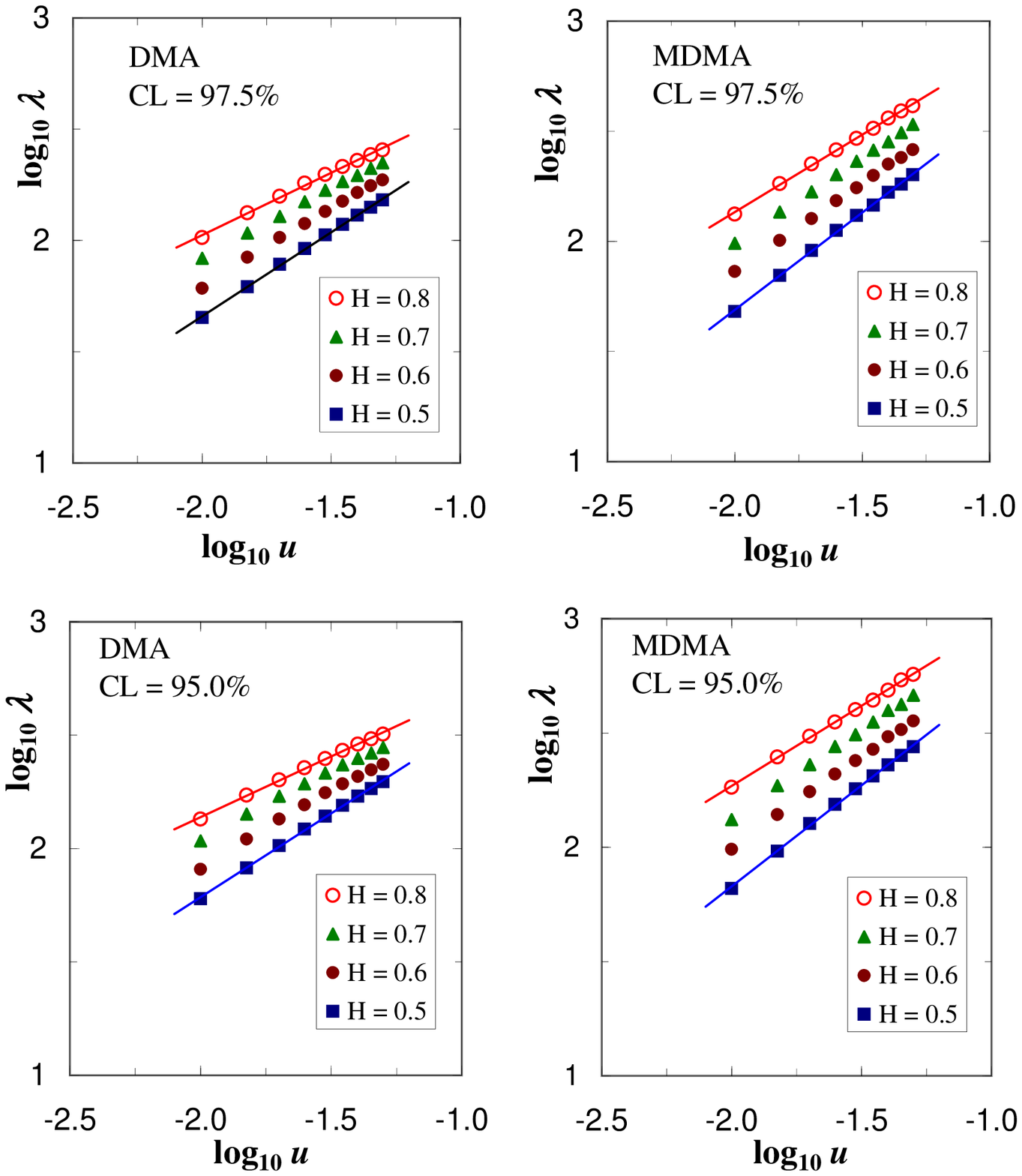,width=15cm}}
\end{center}
\caption{Same as in Fig.\,1 but for dependence of scaling range on $R^2$ ($u=1-R^2$) for signal of length $L=10^3$.}
\end{figure}

\begin{figure}
\begin{center}
{\psfig{file=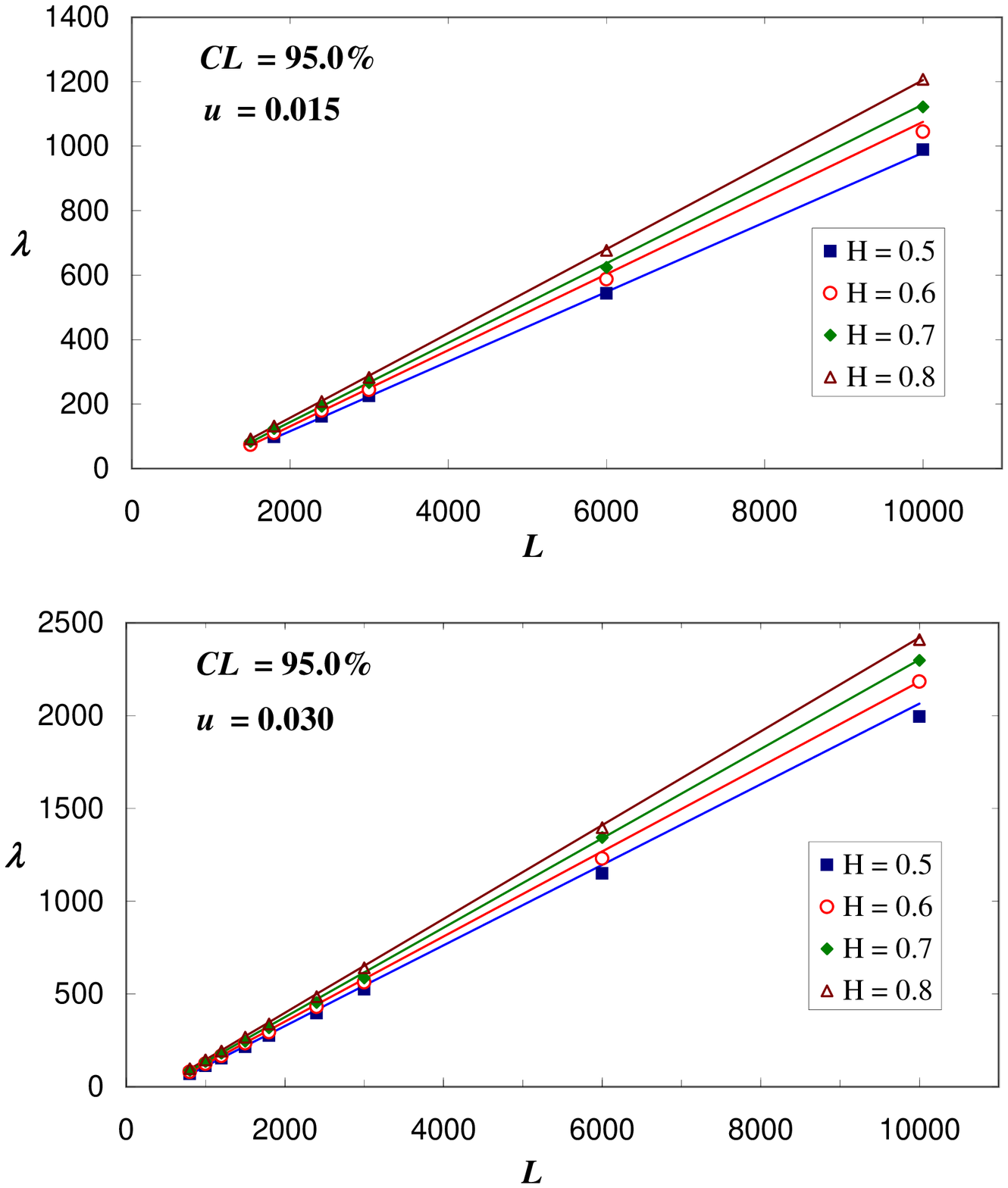,width=15cm}}
\end{center}
\caption{Dependence of scaling range $\lambda$ on data length for DFA for two particular values of $R^2$ at the confidence level $95\%$. The plots look qualitatively the same for wide range of other $u=1-R^2$ (not shown). Perfect linear dependence is observed.}
\end{figure}

\begin{figure}
\begin{center}
{\psfig{file=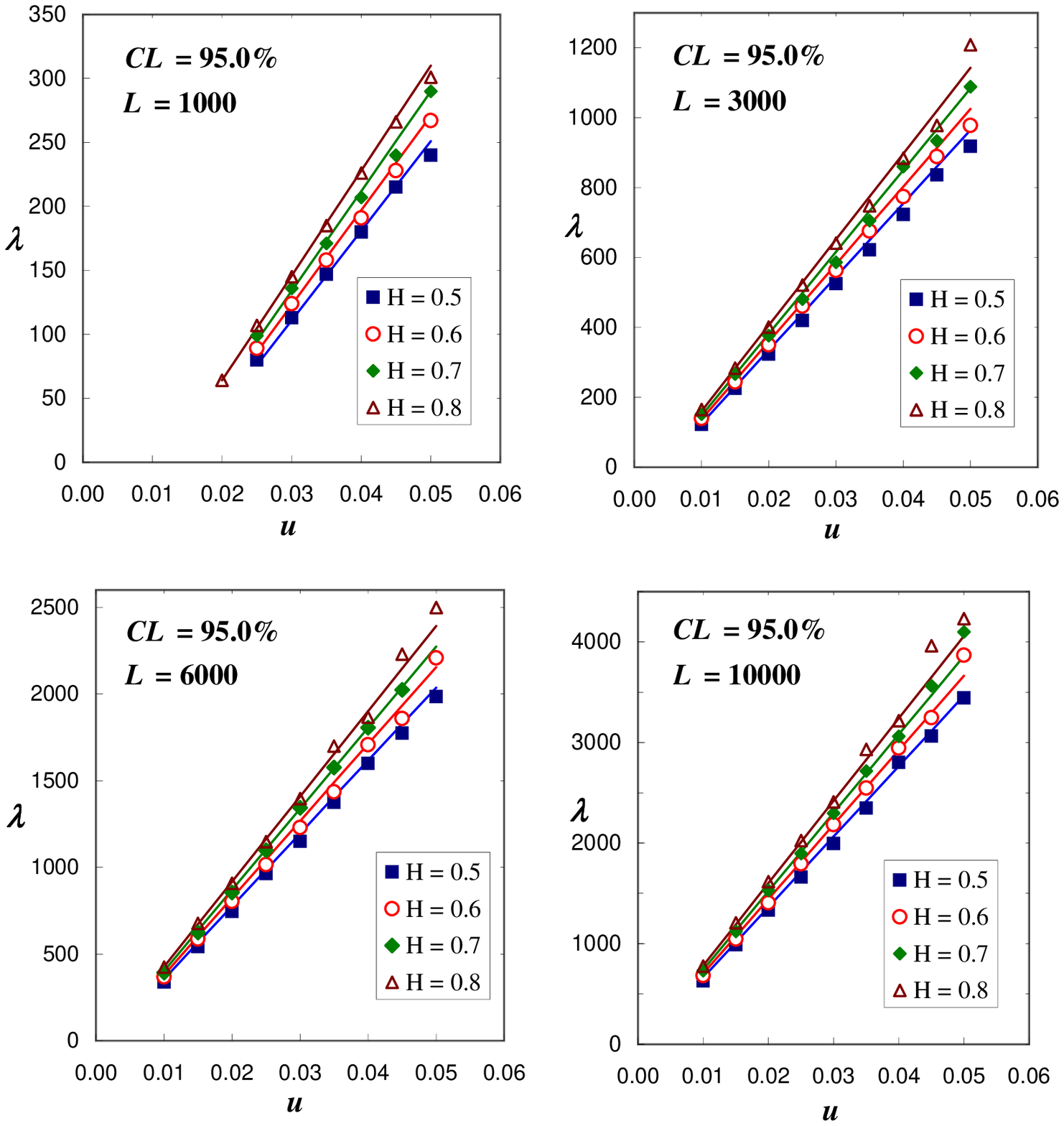,width=15cm}}
\end{center}
\caption{Same as in Fig.\,3 but for dependence on $u=1-R^2$ for various signal lengths.}
\end{figure}

\begin{figure}
\begin{center}
{\psfig{file=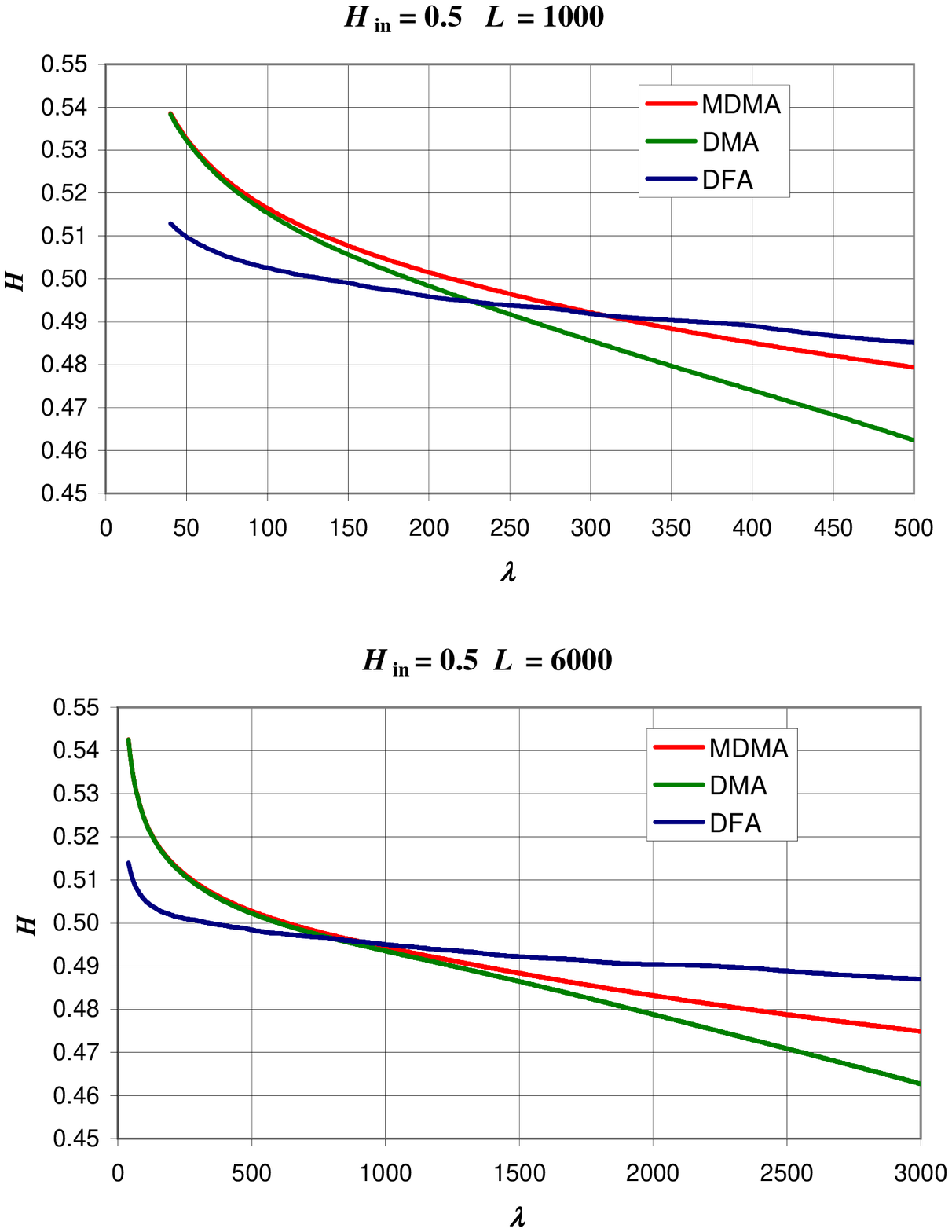,width=15cm}}
\end{center}
\caption{Reproduction of Hurst exponent for uncorrelated data within DFA, DMA and MDMA. The results for two length of data are shown ($L=1000$ and $L=6000$).}
\end{figure}

\begin{figure}
\begin{center}
{\psfig{file=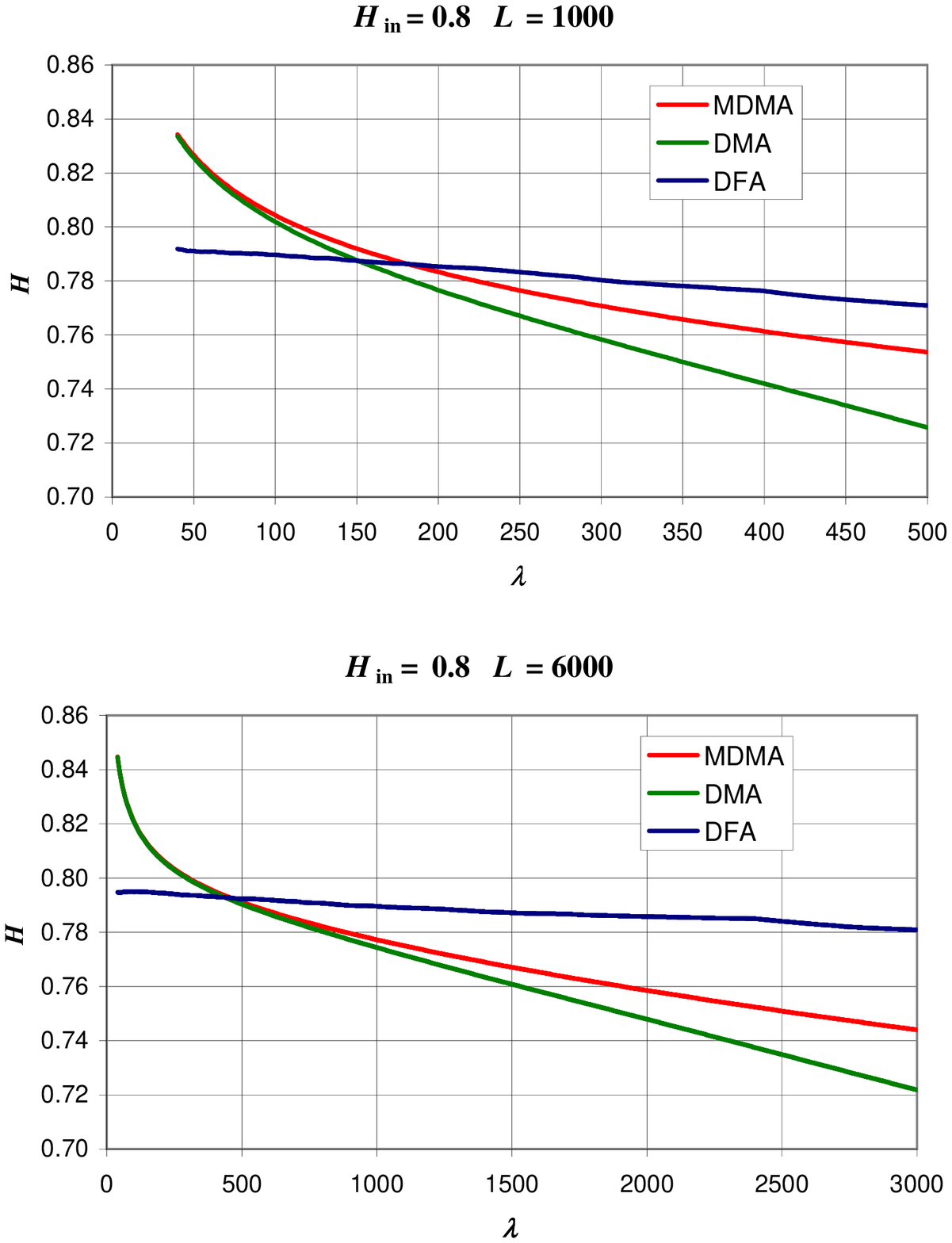,width=15cm}}
\end{center}
\caption{Same as in Fig.\,5 but for persistent synthetic signal with input scaling exponent value $H_{in}=0.8$.}
\end{figure}

\begin{figure}
\begin{center}
{\psfig{file=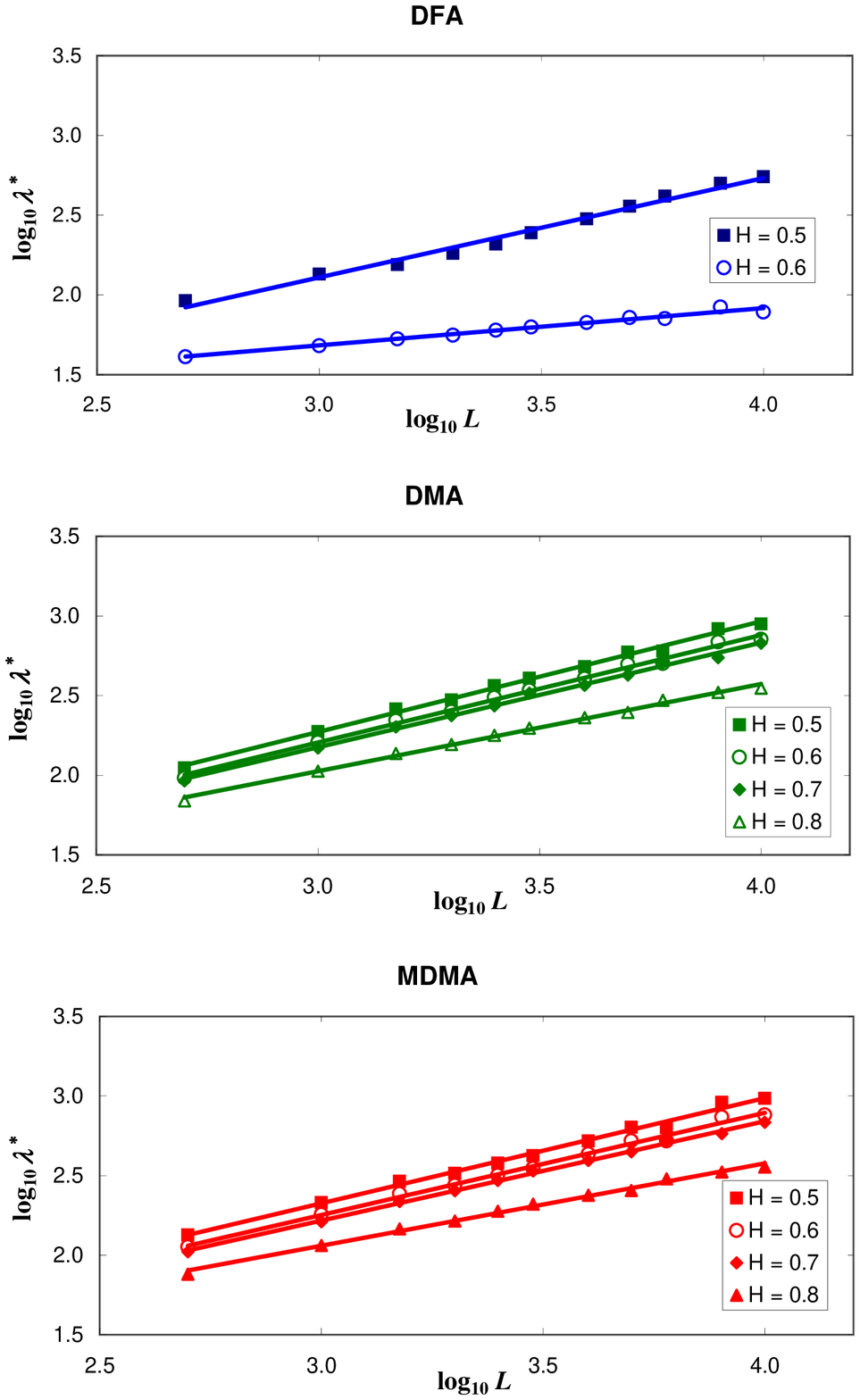,width=11.5cm}}
\end{center}
\caption{Power law dependence between scaling range $\lambda$ well reproducing the true value of $H$ exponent for different kinds of fractional Brownian motion signals with uncorrelated ($H=0.5$), weakly autocorrelated ($H=0.6$) or strongly autocorrelated ($H=0.7,\, 0.8$) increments. DFA, DMA and MDMA results are compared together. The fitting parameters of power law dependence are collected with uncertainties in Table 1.}
\end{figure}

\begin{figure}
\begin{center}
{\psfig{file=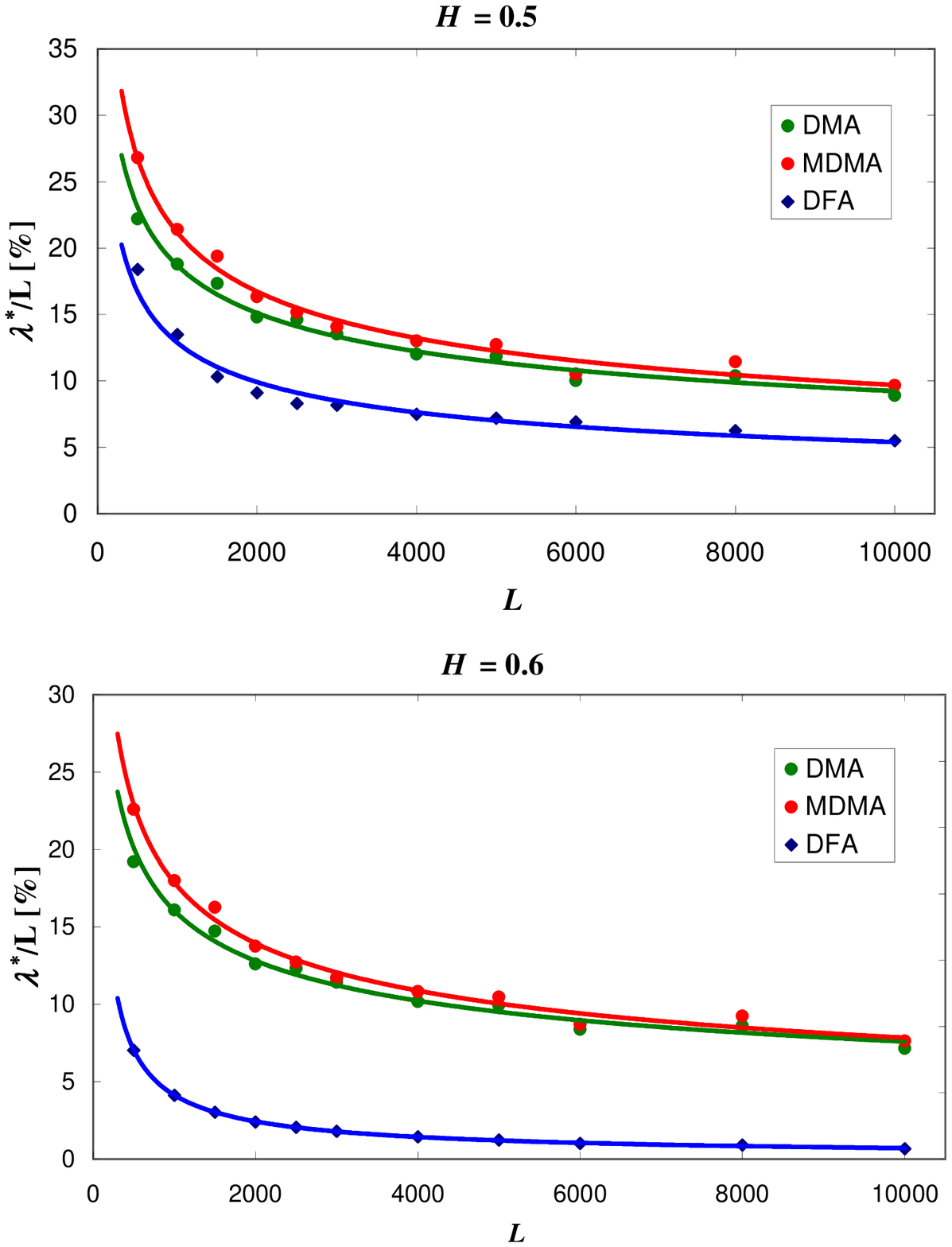,width=15cm}}
\end{center}
\caption{Percentage relative scaling range $\lambda^*/L$ reproducing the input value of $H$ with uncertainty $\delta H /H \lesssim 1\%$. The results for different detrending methods  and for uncorrelated or moderately correlated data are shown.}
\end{figure}

\begin{figure}
\begin{center}
{\psfig{file=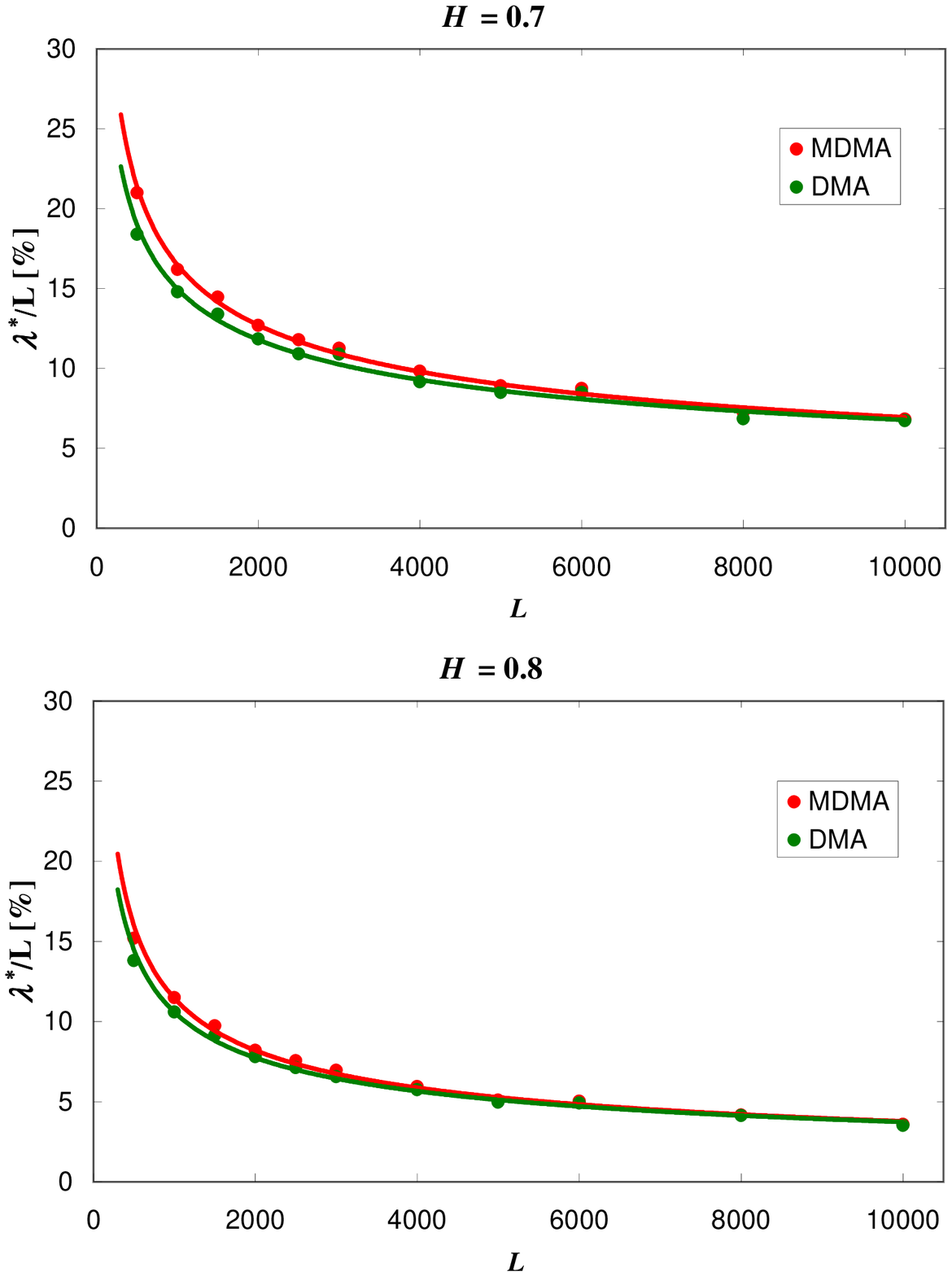,width=15cm}}
\end{center}
\caption{Same as in Fig.\,8 but for more persistant signals ($H=0.7$ and $H=0.8$). Notice that plots for DFA are absent since this method does not warranty solutions here.}
\end{figure}

\end{document}